*Subject Section*

# Insights into performance evaluation of compound-protein interaction prediction methods

Adiba Yaseen[1,*], Imran Amin[2], Naeem Akhter[1], Asa Ben-Hur[3] and Fayyaz Minhas[4*]

[1]Department of Computer and Information Sciences (DCIS), Pakistan Institute of Engineering and Applied Sciences (PIEAS), Islamabad, Pakistan.

[2]National Institute for Biotechnology and Genetic Engineering, Faisalabad, Pakistan.

[3]Department of Computer Science, Colorado State University, Fort Collins, USA.

[4]Tissue Image Analytics Centre, Department of Computer Science, University of Warwick, Coventry, UK.

*To whom correspondence should be addressed.

**Abstract**
**Motivation:**
Machine learning based prediction of compound-protein interactions (CPIs) is important for drug design, screening and repurposing studies and can improve the efficiency and cost-effectiveness of wet lab assays. Despite the publication of many research papers reporting CPI predictors in the recent years, we have observed a number of fundamental issues in experiment design that lead to over optimistic estimates of model performance.
**Results:**
In this paper, we analyze the impact of several important factors affecting generalization performance of CPI predictors that are overlooked in existing work:
1. Similarity between training and test examples in cross-validation
2. The strategy for generating negative examples, in the absence of experimentally verified negative examples.
3. Choice of evaluation protocols and performance metrics and their alignment with real-world use of CPI predictors in screening large compound libraries.

Using both an existing state-of-the-art method (CPI-NN) and a proposed kernel based approach, we have found that assessment of predictive performance of CPI predictors requires careful control over similarity between training and test examples. We also show that random pairing for generating synthetic negative examples for training and performance evaluation results in models with better generalization performance in comparison to more sophisticated strategies used in existing studies. Furthermore, we have found that our kernel based approach, despite its simple design, exceeds the prediction performance of CPI-NN. We have used the proposed model for compound screening of several proteins including SARS-CoV-2 Spike and Human ACE2 proteins and found strong evidence in support of its top hits.
**Availability:** Code and raw experimental results available at https://github.com/adibayaseen/HKRCPI
Contact: Fayyaz.minhas@warwick.ac.uk
**Supplementary information:** Supplementary data files are available as part of the GitHub repository.

# 1 Introduction

Compound Protein Interaction (CPI) prediction is an important task in Target Compound Screening for identifying protein targets of compounds, drug design, and drug repurposing studies (Schirle and Jenkins 2016). Affinity chromatography (Broach and Thorner 1996) and protein microarrays (Lee and Lee 2016; Zhao et al. 2021) are among the most frequently used experimental methods for the identification of CPIs. However, such wet-lab approaches can be expensive and time-taking (W. Zhang, Pei, and Lai 2017) (Paul et al. 2010). The emergence of pandemics such as Ebola and COVID-19 and the global challenge of antimicrobial resistance have highlighted the need of improving efficiency and throughput in drug design (Thafar et al. 2019). Consequently, CPI prediction using computational methods has become an attractive area of research (X. Chen et al. 2016) as such approaches can improve the cost, time, and efficiency of drug discovery in contrast to experimental methods (Mazandu et al. 2018).

## 1.1 Approaches for Compound Protein Interaction Prediction

Conventionally, structure-based and ligand-based virtual screening are the most well-researched areas of drug discovery (Lim et al. 2021). However, these methods require the three-dimensional (3D) structure of the protein of interest. As a consequence, machine learning (ML) based methods that use sequence characteristics of proteins and chemical structural representations of compounds for interaction prediction have been developed (Bredel and Jacoby 2004) (Bleakley and Yamanishi 2009; Gönen 2012; Y. Wang and Zeng 2013). Based on the representation of proteins and compounds used in them, these computational methods can be categorized into three main classes: feature representation-based methods, similarity-based methods, and end-to-end learning methods. Similarity-based methods are based on the assumption that similar drugs tend to target similar proteins and vice versa (R. Chen et al. 2018). In feature representation-based approaches (Ding et al. 2014), features from compounds and proteins are extracted and fed to a machine learning model such as the nearest neighbor predictor, bipartite local models (Bleakley and Yamanishi 2009), Bayesian matrix factorization-based kernels (Gönen 2012), gaussian contact profiling (van Laarhoven, Nabuurs, and Marchiori 2011), pairwise kernel method (Jacob and Vert 2008), etc. Comparative analysis by Ding et al. has shown that PKM outperforms other approaches (Ding et al. 2014).

In recent years, researchers have developed multiple deep learning models for CPI prediction. DeepDTA (Öztürk, Özgür, and Ozkirimli 2018) extracts real-valued sparse feature representations of proteins as well as compounds using convolutional neural networks (CNNs) and appends these features through the final fully connected layer. WideDTA (Öztürk, Ozkirimli, and Özgür 2019) and Conv-DTI (S. Wang et al. 2020) also used an analogous idea with additional features, ligand structural similarity, and information about protein domains and motifs to enhance model accuracy. For representation compound structures, GraphDTA (Nguyen et al. 2021) and CPI–GNN (Tsubaki, Tomii, and Sese 2019) used novel graph neural networks (X.-M. Zhang et al. 2021) (GNNs) as an alternative to CNNs. CPI-NN was shown to outperform other embedding-based methods.

## 1.2 Issues in performance assessment of CPI models

Despite the increased sophistication of CPI models through deep learning, the generalization performance of existing approaches on independent or real-world datasets is still not perfect (Riley 2019). One of the fundamental issues behind this is biased and overly-optimistic performance assessment strategies arising from the use of unsuitable datasets, poor non-redundancy control in train-test data splitting in cross-validation, improper procedures for generation of negative example, lack of independent test sets, and choice of performance metrics. Here, we discuss each of these issues in further detail.

A number of ML-based CPI prediction models have used the MUV (Rohrer and Baumann 2009), DUD-E (Mysinger et al. 2012) and Human-CPI datasets (Tsubaki, Tomii, and Sese 2019; Liu et al. 2015) for model training and performance evaluation. However, these datasets do not contain true or experimentally verified negative examples and may have a large degree of redundancy between proteins and compounds which can lead to biased machine learning models (Lieyang Chen et al. 2019), (Sieg, Flachsenberg, and Rarey 2019) (Lifan Chen et al. 2020).

Another issue associated with the performance assessment of ML CPI models is the protocol used for generating negative examples. As there is no standardized dataset of negative examples for compound-protein interaction prediction, researchers in this domain resort to one of two approaches for the generation of "synthetic" negative examples for training and performance assessment of machine learning models: Random pairing and inter-class similarity-controlled negative example generation. In random pairing, proteins and compounds in the positive set are simply randomly paired for generating synthetic negative examples after exclusion of known positive pairs as in the dataset used in CPI-NN (Tsubaki, Tomii, and Sese 2019). However, researchers have argued that random-pairing can produce examples that are highly similar to positive examples and this can add labeling noise in training (Ding et al. 2014). As a consequence, they have proposed that negative examples should be generated with controls over inter-class similarity. This process first creates a candidate negative set through random pairing of compounds and proteins. Then a similarity function is used to calculate the degree of similarity between a candidate negative example and the given set of positive examples. Only those candidate negative examples are added to the final negative set whose similarity score with positive examples is lower than a pre-specified threshold resulting in negative examples that are sufficiently dissimilar to known positive examples (Ding et al. 2014). However, as in the case of protein-protein interaction prediction models (Ben-Hur and Noble 2006), the use of similarity-controlled negative example generation in model evaluation can result in overly optimistic performance results with a high likelihood of generalization failure on real-world test sets.

A number of existing approaches also use an equal number of positive and negative examples even though the number of compounds that can be expected to bind to a given protein can be significantly smaller in comparison to the size of the universe of possible compounds. This results in the generation of a large number of false positives in real-world applications. Furthermore cross-validation protocols employed in most existing ML CPI models also do not consider protein sequence and compound similarity in generating training and test folds resulting in overly optimistic performance estimates as the training set can contain examples that are very similar to test examples. Ideally, the examples in the test folds should be sufficiently different from training examples to reflect real-world use cases.

Lastly, existing methods report areas under the Receiver Operating Characteristic or Precision-Recall curves (abbreviated as AUCROC and AUCPR) as performance metrics. However, given that such approaches are typically used for screening interactions from a large number of candidate compound-protein pairs for wet-lab validation, these metrics may not provide a directly interpretable estimate of how good a method is at ranking interacting compounds of a protein.

**Table 1** Selection criteria applied to Binding DB for generating the negative dataset

| Selection Criteria | Total examples |
|---|---|
| Number of Compound-Protein Pairs in Binding-DB (June 2021) | 22,782,226 |
| Examples with single-chain protein targets | 2,169,607 |
| Examples with Binding Affinity measured in terms of $K_i$ | 490,605 |
| Examples with valid Uniport identifiers | 487,732 |
| Examples with valid SMILES strings | 485,550 |
| Examples with valid Inhibition constant $K_i$ values | 479,802 |
| Compound-protein pairs with $K_i \geq 1783000\ nM$ | 4,355 |
| **Final Negative set after removal of duplicate examples** | **3,657** |

### 1.3 Contributions of this work

In this work, we highlight the issues discussed above with a number of experiments using an existing CPI prediction model (CPI-NN) (Tsubaki, Tomii, and Sese 2019) as well as a novel heterogeneous kernel-based approach. We suggest improvements in the evaluation protocol used for performance assessment of such models in terms of negative example generation as well as performance metrics. We report the prediction results of the proposed approach for screening candidate compounds for a number of test proteins not included in the data sets used in model construction including SARS-CoV-2 Spike protein and Human ACE2.

## 2 Methods

In this section, we discuss details of our datasets, experiments and machine learning method design for compound protein interaction prediction.

### 2.1 Datasets

#### 2.1.1 .Liu et al. Human CPI Dataset (HCPI)
We use the human protein-compound interaction dataset originally proposed by (Liu et al. 2015) and employed in a number of existing methods such as CPI-NN (Tsubaki, Tomii, and Sese 2019). In this dataset, positive examples consisting of protein-compound pairs were collected from two experimentally verified databases: DrugBank 4.1 (Wishart et al. 2008) and Matador (Günther et al. 2008). This dataset has 3,364 positive examples of interacting protein-compound pairs constituting 852 unique proteins and 1179 unique compounds. It also contains an equal number of negative examples obtained by randomly pairing proteins and compounds in the positive set provided as part of the CPI-NN code repository (Tsubaki, Tomii, and Sese 2019).

#### 2.1.2 Non-redundant Human CPI dataset (NR-HCPI)
We found that the aforementioned HCPI dataset by Liu et al. and used in CPI-NN (Tsubaki, Tomii, and Sese 2019) contains duplicated examples which can lead to an overestimation of prediction performance.

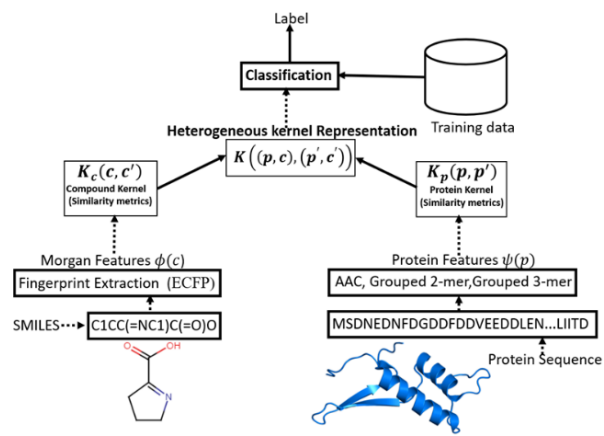

**Figure 1** Concept diagram of the proposed Kernel Compound-Protein Interaction (Kernel-CPI) Prediction. Protein sequence and SMILES are given as input, amino acid composition and grouped k-mer (k = 2, k = 3) features are extracted from the protein sequence and concatenated into a single feature vector $\psi(p)$ for computing a protein-level kernel $K_p(p, p')$. The Morgan Fingerprint $\phi(c)$ is extracted from SMILES representation of a compound to compute the compound kernel $K_c(c, c')$. The two kernels are then combined into a joint heterogeneous kernel representation $K\big((p, c), (p', c')\big)$ for CPI prediction.

We removed these duplicated examples from the HCPI positive set resulting in 2,633 unique positive examples that constitute our NR-HCPI dataset together with negative examples obtained by randomly paired proteins and compounds in the positive set excluding any pairs already included in the positive set. We generated different ratios of positive to negative examples (P:N = 1:1, 1:3, 1:5, and 1:7) for the evaluation of predictive performance under more realistic evaluation scenarios with high-class imbalance. In conjunction with this dataset, we also utilized a non-redundant cross-validation (NRCV) protocol which is detailed in the performance evaluation section.

#### 2.1.3 Binding DB dataset

As discussed in the Introduction, one of the fundamental issues with protein-compound interaction datasets is the lack of experimentally verified negative examples. For performance assessment of CPI prediction methods and for studying the impact of various approaches for generating synthetic negative examples, we have used the binding affinity values of protein-compound pairs in the latest version (June 2021) of Binding DB (Gilson et al. 2016) with a total of 22,782,226 examples. For this purpose, we applied a number of data filtering steps (shown in Table-1) such as using only single-chain protein targets with experimentally verified inhibition constant values that are sufficiently high to ensure very low probability of interaction (Ganesh et al. 2005; Çavdar et al. 2012) to select our final dataset of 3,657 negative examples.

#### 2.1.4 Superdrug bank for drug repurposing

For drug repurposing analysis, we use the SuperDRUG2 (version 2) database (Siramshetty et al. 2018) of approved and commercially available drugs with a total of 3,633 unique small molecules. We have also used the Superdrug bank molecules for screening potential targets of SARS-Cov-2 Spike protein and the human ACE2 protein.

### 2.2 Machine Learning Models

For performance analysis, we have used the existing CPI-NN method which gives state-of-the-art results over the same datasets (Tsubaki, Tomii, and Sese 2019). CPI-NN has been validated for human and C.

elegans proteins with high AUC-ROC (0.95 and above) and under different class ratio settings. We used the publicly available code of CPI-NN and conducted experiments with various cross-validation and assessment strategies ourselves after verifying the reproducibility of the results using the same experimental settings as reported in the original CPI-NN paper.

We have also developed a simple kernel-based approach for CPI prediction (see Fig. 1). For this purpose, we model compound protein interaction (CPI) prediction as a classification problem in which every example $x \equiv (c, p)$ consists of a protein $p$ and a compound $c$ with corresponding feature representations $\psi(p)$ and $\phi(c)$, respectively. Each example in the training dataset $D = \{((p_i, c_i), y_i) | i = 1 \ldots N\}$ is associated with a binary label $y_i \in \{-1, +1\}$ indicating whether the corresponding protein and compound interact $(+1)$ or not $(-1)$. Features are extracted from protein sequence as well as from SMILES of compounds for heterogeneous modeling of the CPI problem as discussed below.

### 2.2.1 Protein Features

*Amino Acid Composition (AAC)*

In order to capture amino-acid specific binding characteristics of proteins with their target compounds in the predictive model, we have used the amino acid composition (AAC) of protein (denoted by $\psi_{AAC}(p)$) which is a 20-dimensional vector representation of a protein sequence containing the frequency of occurrence of various amino acids in the protein sequence (K. Huang et al. 2020).

*Grouped k-mer Features*

In order to model the physiochemical similarity across amino acids, we used grouped k-mer composition of proteins as a feature vector. In this approach, each amino acid in a protein is assigned one of seven predetermined groups based on its physicochemical characteristics (Hashemifar et al. 2018) and the counts of all possible group-level k-mers are used as a feature vector. For $k = 2$ and $k = 3$, this results in $7^2 = 49$- and $7^3 = 343$-dimensional features of a protein denoted by $\psi_2(p)$ and as $\psi_3(p)$, respectively.

### 2.2.2 Compound Features

For modeling compound characteristics, we extract features from SMILES of compounds in the compound protein interaction pair using its Extended-Connectivity Fingerprint (ECFP) (also known as Morgan Fingerprint) (M. Veselinovic et al. 2015) using RDKit (Cao et al. 2013). This fingerprint is a topological feature of a chemical compound and captures its structural confirmation within a given radius. The feature dimension of this representation is 1024 for a radius of 3 atoms.

### 2.2.3 Heterogeneous Kernel Representation

We have developed a heterogeneous feature space kernel representation for compound-protein interaction prediction. As each classification example in this problem comprises a protein and compound, we first construct non-linear kernel representations of proteins and compounds separately which are then merged to form a heterogeneous feature space kernel for classification as shown in Figure 1.

*Compound Similarity kernel*

We use the compound feature representation $\phi(c)$ to construct a radial basis function (RBF) similarity kernel between pairs of compounds as follows:

$$K_c(c_i, c_j) = exp(-\gamma_c \|\phi(c_i) - \phi(c_j)\|^2)$$

In this equation $\phi(c_i)$ and $\phi(c_j)$ are Morgan Fingerprint feature vectors as described in the previous section. The kernel $K_c(c_i, c_j)$ essentially measures the degree of similarity of two compounds in the feature space in a non-linear manner with a single hyper-parameter $\gamma_c > 0$.

*Protein Similarity kernel*

For a protein $p$, all three feature vectors of protein sequence $p$ are concatenated in a feature representation $\psi(p)$ resulting in a 412-dimensional column vector of the protein features as shown below.

$$\psi(p) = [\,\psi_{AAC}(p)\,|\,\psi_2(p)\,|\,\psi_3(p)\,]$$

This feature representation is then used to generate a protein-protein similarity kernel as follows:

$$K_p(p_i, p_j) = exp(-\gamma_p \|\psi(p_i) - \psi(p_j)\|^2)$$

*Heterogeneous kernel Representation and Classification*

Based on the protein and compound similarity kernels, we construct a heterogeneous feature space kernel between pairs of examples $(p, c)$ and $(p', c')$ each consisting of a protein and a compound as follows:

$$K((p, c), (p', c')) = \langle \varphi(p, c), \varphi(p', c') \rangle = \left(K_p(p, p') + K_c(c, c')\right)^2$$
$$= K_p(p, p')^2 + K_c(c, c')^2 + 2K_p(p, p')K_c(c, c')$$

This joint kernel essentially measures the degree of similarity between pairs of examples with each example being a protein-compound pair. Note that the joint kernel is a quadratic sum of the protein and compound kernels which gives rise to an abstract and nonlinear joint feature space $\varphi(p, c)$ for compound-protein pairs with the kernel $K$ being an implicit generalized dot product between $\varphi(p, c)$ and $\varphi(p', c')$. The product $K_p(p, p')K_c(c, c')$ in the above formulation implicitly corresponds to the tensor-product of the protein and compound feature spaces. It is also important to note that two examples will have a high kernel score if the corresponding proteins and compounds in the two examples are similar. The joint kernel over the training dataset $D = \{((p_i, c_i), y_i) | i = 1 \ldots N\}$ is used for training a kernelized Support Vector Machine (SVM) (Vapnik and Izmailov 2017) which is then used to infer the prediction score $f(p, c)$ for a given test example $(p, c)$. This approach is in line with the work by (Jacob and Vert 2008) with major differences in the choice of constituent kernels and construction of the joint kernel (see supplementary material for comparative results).

## 3 Performance Comparison & Screening Experiments

We have designed multiple experiments to identify issues in performance evaluation and generalization of CPI predictors which are described in this section.

### 3.1 Five-fold stratified cross-validation (CV)

For direct comparison with previous methods, we have used stratified five-fold cross-validation which is typically used for reporting CPI prediction results. In five-fold stratified cross-validation, the dataset is divided into 5 equal folds such that the ratio of examples of every class in each fold remains the same as the overall class proportion. Each cross-validation experiment is repeated ten times to obtain the average and standard deviation of different performance metrics such as AUCROC and AUC-PR.

### 3.2 Non-Redundant Five-Fold Cross-Validation (NRCV)

One of the limitations of five-fold cross-validation is that very similar proteins or compounds may end up in different folds resulting in an overly optimistic assessment of prediction performance. To estimate the generalization performance in a real-world setting where test proteins may not share very high sequence similarity with proteins in the training set, we have also performed a more stringent non-redundant cross-validation analysis which has not been performed in previous studies. For this purpose, proteins in the NR-HCPI dataset are first clustered based on a given sequence identity threshold through CD-HIT (Y. Huang et al. 2010). These clusters are then divided into five folds such that no two folds have examples from the same cluster while ensuring that the number of examples in every fold remains approximately the same. This guarantees that the sequence similarity of proteins in examples in a test fold is always less than a specified threshold with proteins in the training set. We used two different sequence similarity thresholds (40% and 90%) in our analysis.

### 3.2.1 Validation over experimentally verified negative examples from Binding DB

In addition to classical 5-fold cross-validation and non-redundant cross-validation, we have also analyzed the prediction quality of our proposed method as well as CPI-NN on an external set containing experimentally verified negative examples from Binding DB as described in the dataset section. In this experiment, the ML model is trained on four folds of non-redundant cross-validation as described above. However, the original negative examples in the test fold are replaced with experimentally verified negative examples from Binding DB. This process is repeated by alternating across different folds and then multiple runs to generate mean and standard deviation values of performance metrics.

### 3.2.2 Analysis of Negative Example Generation Strategies

As discussed in the Introduction section, there are two strategies used in the literature for generating negative examples: Random Pairing and Similarity Controlled Negative Example Generation. In this work, we systematically compare these strategies for training and performance assessment of the proposed model. For this purpose, we have developed the algorithm shown in Table-2 to generate negative examples at different inter-class similarity thresholds using kernel-based calculations. This algorithm can be used to generate a desired number of synthetic negative examples by controlling their degree of similarity to examples in a given positive set based on an inter-class similarity threshold $\alpha \in [0,1]$. For our systematic comparison, we first pick a value of α and then generate synthetic negative examples through this algorithm for training and performance evaluation. It is important to note that for sufficiently high values of α, i.e., α→1, this algorithm essentially generates randomly-paired negative examples which can be similar to known positive examples whereas for low values (α→0), the generated negative examples are completely dissimilar to known positive examples. The resulting data of positive and synthetic negative examples is then divided into five folds in a stratified manner as for non-redundant cross-validation. Similar to NRCV, training is performed on four folds followed by testing on examples in the held-out set in two different ways: first by using the held-out set of positive and synthetic negative examples and, secondly, by using the held-out set of positive examples and "true" negative examples from Binding DB. The process is then repeated for different values of α. Differences in the predictive performance of a given method between the cross-validation protocol and the

**Table 2** Algorithm for generation of negative examples with desired inter-class similarity threshold

**Inputs:**
Set of positive examples $\wp = \{(p_i, c_i) | i = 1 \ldots P\}$
Set of unique proteins $P_U$ ($P_U = \{p | (p,c) \in \wp\}$)
Set of unique compounds $C_U$ ($C_U = \{c | (p,c) \in \wp\}$)
Desired number of negative examples N (based on P:N ratio)
Desired inter-class similarity threshold $\alpha \in [0,1]$
**Output:** Set $\aleph$ of N negative examples with similarity to positive examples $\wp$ less than $\alpha$
**Algorithm**:
Initialize $\aleph \leftarrow \{\}$
While $|\aleph| < N$:
 Randomly select a protein-compound pair $(p,c)$ from $P_U \times C_U$
 Calculate similarity of candidate negative example with positive set as follows:
 $\alpha_{pc} = max_{p' \in P_U - \{p\}} K_p(p, p') max_{c' \in C_U - \{c\}} K_c(c, c')$
 If $\alpha_{pc} < \alpha$ and $(p,c) \notin \wp \cup \aleph$: $\aleph \leftarrow \aleph \cup \{(p,c)\}$
Return $\aleph$

evaluation with true negative examples from Binding DB indicate systematic biases resulting from synthetic negative example generation strategies.

### 3.2.3 All-vs-All Target Compound Screening

In a practical setting, compound-protein interaction prediction approaches are used for screening a large number of compounds for potential binding with a target protein of interest. Ideally, interacting compounds of a given protein should rank close to the top in comparison to non-interacting compounds in the screening library based on prediction scores of all test examples from the predictor.

However, cross-validation experiments used in previous works do not model this "screening" use case as they are restricted to a fixed evaluation data set and do not analyze how a predictor would rank known interacting partners in a setting in which all compounds are paired with all proteins. In this work, we have performed in silico screening of all unique compounds against all proteins in a given test set. This all-vs-all pairwise screening is useful for drug discovery and repurposing studies and is carried out by computing the prediction score of all possible pairs of proteins and compounds in a test set using a prediction model and calculating how often a predictor ranks a known interacting pair in its top predictions. We have performed three different screening experiments for comparison between CPINN and the proposed model:

*Screening with Non-redundant Cross-validation (NRCV)*
In this experiment, we train a model using training folds of the NRCV dataset and then compute prediction scores of all-vs-all compound-protein pairs in the test fold using the trained model (see supplementary information file for an illustration of the experimental setup). This process is repeated for all five folds of the dataset to compute a rank-based performance metric (RFPP) described in the next subsection.

*Screening SuperDRUG2*
For drug-repurposing analysis with the proposed model, we used the SuperDRUG2 dataset containing 3,633 FDA-approved drugs. In this experiment, a CPI model is first trained on all examples in training folds of the NRCV dataset and then used to generate prediction scores for all proteins in the test fold paired with all compounds in the SuperDRUG2 database (see supplementary material on GitHub for an illustration of the experimental setup). These scores are used to rank known interacting compounds of each protein in the test set relative to the compounds in SuperDRUG2 to compare the predictive performance of CPINN and the proposed model and identify putative compounds in SuperDRUG2 that can bind test proteins in the NRCV dataset.

**Table 3** Results of the proposed approach and CPI-NN over HCPI dataset, non-redundant HCPI dataset and non-redundant cross-validation (NRCV) with mean and standard deviation (in brackets) of AUCs (expressed as percentages). Numbers in parentheses in data show number of examples.

| Data | CPI-NN | | Proposed (Kernel-CPI) | |
|---|---|---|---|---|
| | AUC-ROC | AUC-PR | AUC-ROC | AUC-PR |
| HCPI (5994) | 94.41± (1.19) | 94.01 (2.21) | **98.98 (0.14)** | **99.03 (0.16)** |
| NR-HCPI (5266) | 93.1 (1.06) | 91.44 (0.64) | **93.84 (2.35)** | **94.56 (1.31)** |
| NR-HCPI (NRCV) (5266) | 62.58 (1.1) | 72.52 (5.2) | **69.98 (5.7)** | **77.3 (1.44)** |

*Screening for SARS-CoV-2 associated proteins*

We have also used the proposed model trained over the NR-HCPI dataset to generate predictions for interactions of SARS-CoV-2 Spike protein and Human ACE2 protein across all compounds in SuperDrug2 to identify putative interactions (Goulter et al. 2004), (Xia and Lazartigues 2008; Zou et al. 2020). We then performed a literature search for any experimental evidence of interaction of the top-scoring compound with these proteins or their association with SARS-CoV-2 treatment effects. For this purpose, the proposed model was trained over positive examples in the HCPI dataset after removing lightweight molecules (with molecular weight less than 100) as well as these proteins and using a P: N ratio of 1:7.

#### 3.2.4 Using Ranks for Performance Evaluation

For quantifying the prediction quality of CPI predictors in screening experiments, we have developed an interpretable performance metric called Rank of First Positive Prediction (RFPP) inspired from our previous work on protein-protein interactions (Minhas, Geiss, and Ben-Hur 2014). For a given protein in the test set, RFPP is obtained by first pairing all possible test compounds with the protein and computing the prediction scores of all such examples using the CPI model under evaluation. Then the rank of the highest-scoring compound that is a known interacting partner of the test protein is used as the RFPP value of that protein (see supplementary material for an illustration of this experimental setup). Note that for an ideal predictor, the RFPP for all test proteins should be one, i.e., the top-ranked compound of each test protein should be an interaction partner of that protein.

In order to quantify the predictive quality of a CPI model across all test proteins, we first compute RFPP for all test proteins and then calculate percentiles of the RFPP values across all proteins. The percentile values across all proteins can be used to compare the predictive performance of screening models based on their ability to rank putative compound-protein interactions. The $r^{th}$ percentile of RFPP of a predictor will be $q$ (denoted as $RFPP(r) = q$) if $r\%$ test proteins have at least one known interacting compound in the top q predictions from the predictor. It essentially shows the expected number of compounds that will need to be screened to identify a true interacting partner in wet lab experiments. For an ideal predictor, the RFPP value for all proteins in the test set should be one, i.e., $RFPP(100)=1$. We have generated the RFPP percentile plots of the proposed method as well as CPI-NN. As a baseline we have also plotted the RFPP percentiles of a random predictor which generates random prediction scores given a protein and compound. These values provide more directly interpretable estimates of prediction quality for such screening experiments.

**Table 4** Performance analysis over true negative examples from binding-DB for different training class ratios (P: N). Bold values indicate the best mean AUC percentage ± standard deviation.

| Method | P : N | AUC-ROC | AUC-PR |
|---|---|---|---|
| CPI-NN | 1:1 | 76.81 ± 9.8 | 47.48 ± 5.14 |
| Proposed (Kernel-CPI) | 1:1 | 89.88 ± 2.59 | 72.0 ± 2.02 |
| | 1:3 | 91.19 ± 3.56 | 84.3 ± 4.01 |
| | 1:5 | **91.74 ± 3.35** | 88.34 ± 3.09 |
| | 1:7 | 91.15 ± 2.15 | **88.96 ± 1.86** |

## 4 Results and Discussion

### 4.1 Non-redundant cross-validation analysis is essential for realistic performance evaluation

Previous approaches have used five-fold cross-validation (CV) for performance evaluation over the Human CPI dataset. In order to provide a direct comparison with previously published methods, we have performed five-fold cross-validation with the proposed approach with different class ratios (see Table-3).

The proposed model gives an AUC-ROC and AUC-PR of 99% which is better than CPI-NN (94%). However, as discussed in the introduction section, this high predictive performance result of both CPI-NN and the proposed approach can be attributed to the duplication of examples in the dataset as well as similarity between examples across cross-validation folds. If the duplicated examples are removed, we observe a minor drop in prediction performance of both approaches. In order to get a more realistic estimate of the generalization performance of these methods, we have performed non-redundant cross-validation analysis as discussed in the previous section. Table 3 presents detailed results of this non-redundancy analysis for different ratios of positive to negative (P:N) examples ratios and classifiers at 90% sequence identity thresholds. As expected, the predictive performance of the predictors decreases significantly with the removal of non-redundancy between training and test sets through NRCV. These experiments clearly show that it is very important to analyze prediction performance under non-redundant cross-validation. Results at 40% thresholds are reported in the supplementary material (on GitHub) and follow a similar trend.

### 4.2 Validation over true negative examples from Binding DB allows realistic performance evaluation

As outlined in section 2.3.4, we have used a set of experimentally verified negative examples from the Binding-DB dataset to analyze the prediction performance of the proposed model as well as CPI-NN. For this purpose, both models were first trained on the NR-HCPI dataset with a balanced class ratio. The results of this analysis are given in Table 4 which shows that upon using true negative examples from Binding-DB in testing, the prediction performance of the proposed model is superior to CPI-NN (AUC-ROC of 76.8% vs. 89.9%) which supports the findings from non-redundant cross-validation above. As expected, increasing the ratio of negative examples in training for the proposed method improves the prediction performance over the binding DB test set further.

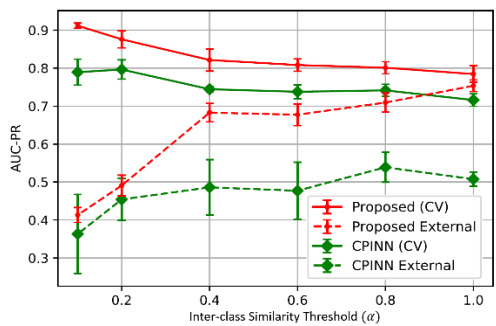
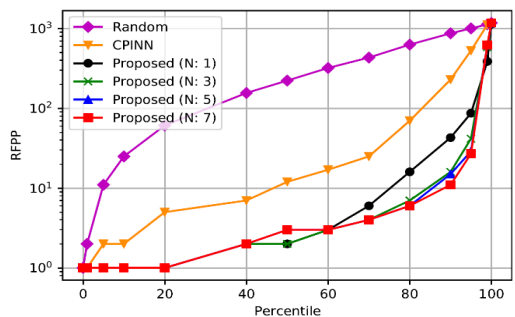

Figure 2 Analysis of the impact of negative example generation strategies. AUC-PR of cross-validation (CV) for CPI-NN and the proposed model shown in solid lines for different values of the inter-class similarity threshold $\alpha$. The results of evaluation over the external dataset are shown in dashed lines. We can see that, as expected, AUCPR of cross-validation (CV) increases with decrease in inter-class similarity of training and test examples for both models. However, evaluation of both models on experimentally verified negative examples in the external set shows that training over highly dissimilar negative examples reduces generalization performance.

### 4.3 Random pairing for generating negative examples yields more realistic and better generalization performance

We have analyzed the impact of different ways of generating synthetic negative examples (random-pairing vs. inter-class similarity controlled negative example generation) on estimation of prediction quality of a CPI model through cross-validation and its generalization performance on an external dataset containing experimentally verified negative examples from Binding-DB.

For this purpose, we have used a procedure that allows us to generate synthetic negative examples by controlling their degree of similarity with a given positive set through an inter-class similarity threshold α (see Section 2.4.5 for complete details of experimental setup). The AUCPR values of both CPINN and the proposed model for cross-validation and the external test set for different values of α is plotted in the Fig 2. It shows that, as expected, if the similarity between the generated synthetic negative examples and the positive set is increased, the AUC-PR values of both CPINN and the proposed approach obtained from cross-validation decrease. This is inline with the findings by (Ding et al. 2014) and support similarity controlled generation of negative examples. However, if models trained over negative examples that are significantly different from the given positive are tested on an external set containing experimentally verified negative examples, their generalization performance is low. This is indicated in Fig 2 by increase in AUCPR values of both CPINN and the proposed approach over the external test set as the value of α is increased. This experiment clearly shows that using random pairing of proteins and compounds is a superior strategy for generating synthetic negative examples as it not only gives more realistic estimates of predictive quality but can improve the performance of CPI models over unseen test sets.

### 4.4 RFPP offers interpretable performance evaluation in screening experiments

Fig 3 shows the RFPP percentiles across all proteins resulting from the all-vs-all screening experiment over the NR-HCPI dataset with non-redundant cross-validation detailed in section 2.4.6. In this experiment, a CPI model is first trained over examples in the training folds of the non-redundant cross-validation dataset and then used to rank all possible pairs of proteins and compounds in the test set to see how good is the method at ranking known interacting compounds for all proteins through the RFPP metric. The total number of all such possible combinations in this dataset is ~292,500. It shows that for 85% of test proteins, the proposed approach is able to find at least one known interacting compound of those proteins in its top 10 hits (i.e., RFPP(85) = 10) whereas, for CPINN, only 50% proteins have at least one known hit in its top 10 predictions for each protein (RFPP(50) = 10). In contrast, a random predictor can be expected to have at least one interacting compound in its top 10 predictions for only 5% of proteins in this test set. This clearly shows the efficacy of the proposed approach as well as the ease of interpreting results of model evaluation through the use of RFPP in screening experiments. As expected, we also see that adding more randomly paired negative examples to training improves RFPP further.

Figure 3 Percentiles of RFPP across proteins in the NRCV screening experiment.

### 4.5 Target Compound Screening (TCS) over for drug repurposing using SuperDRUG2

In order to evaluate the prediction performance of the proposed model and CPINN for possible drug-repurposing studies, we have conducted a virtual screening experiment using the FDA approved drugs in the SuperDRUG2 dataset. For this purpose, we score all possible (~ 908,250) pairs of proteins from the NR-HCPI with compounds from SuperDRUG2. All these predictions from the proposed model are made available to the community as supplementary results. As an additional step, we have also calculate the RFPP percentiles across all proteins from the proposed model as well as CPINN for this screening experiment which are given in the supplementary file. For a random predictor, we can expect to find at least one true interacting compound in the top 10 hits for only 3% of the proteins in this analysis. However, CPINN and the proposed model are able to identify at least one interacting compound for 50% and 75% of proteins, respectively.

### 4.6 Target Compound Screening for Human ACE2 and SARS-CoV-2 Spike Proteins

The results of *in silico* screening of compounds in the SuperDRUG2 dataset for Human ACE2 (Uniprot ID: Q9BYF1) and SARS-Cov-2 Spike (Uniprot ID: P59594) proteins through the proposed method are given in the supplementary file (on GitHub) which shows the top 100 predictions of our model for ACE2 and Spike protein along with evidence from the literature supporting the predicted interaction. We have found that the proposed model is able to identify a number of compounds as potential interaction partners of these proteins even though these were not included in its training. Specifically, we have identified Trandolapril, Dimethyl Sulfoxide (DMSO), Remdesivir, Ramipril, N-Acetylglucosamine, Perindopril, Sunitinib and Glutathione in our top hits for ACE2 binding with strong support from experiments and in silico studies in the literature. Similarly, N-Acetylglucosamine, DMSO, Remdesivir, Sunitinib, Nilotinib, Dasatinib and Sorafenib show binding potential with the spike protein of SARS-Cov-2 with strong support in recent literature.

## Conclusions

In this work, we have identified a number of shortcomings in experiment design approaches for CPI prediction. We hope that this work will enable the community to address these issues so that the future CPI models are more effective in prediction of protein-compound interactions for novel cases.

## Acknowledgements and Funding

This work is supported by HEC NRPU 6085. FM is also support by the PathLAKE consortium grant at University of Warwick.

## References

Baysal, Ömür, Naeem Abdul Ghafoor, Ragıp Soner Silme, Alexander N. Ignatov, and Volha Kniazeva. 2021. "Molecular Dynamics Analysis of N-Acetyl-D-Glucosamine against Specific SARS-CoV-2's Pathogenicity Factors." PloS One 16 (5): e0252571.

Ben-Hur, Asa, and William Stafford Noble. 2006. "Choosing Negative Examples for the Prediction of Protein-Protein Interactions." BMC Bioinformatics 7 (1): S2. https://doi.org/10.1186/1471-2105-7-S1-S2.

Bleakley, Kevin, and Yoshihiro Yamanishi. 2009. "Supervised Prediction of Drug–Target Interactions Using Bipartite Local Models." Bioinformatics 25 (18): 2397–2403. https://doi.org/10.1093/bioinformatics/btp433.

Bredel, Markus, and Edgar Jacoby. 2004. "Chemogenomics: An Emerging Strategy for Rapid Target and Drug Discovery." Nature Reviews Genetics 5 (4): 262–75. https://doi.org/10.1038/nrg1317.

Broach, J. R., and J. Thorner. 1996. "High-Throughput Screening for Drug Discovery." Nature 384 (6604 Suppl): 14–16. https://doi.org/10.1038/384014a0.

Cagno, Valeria, Gaelle Magliocco, Caroline Tapparel, and Youssef Daali. 2021. "The Tyrosine Kinase Inhibitor Nilotinib Inhibits SARS-CoV-2 in Vitro." Basic & Clinical Pharmacology & Toxicology 128 (4): 621–24. https://doi.org/10.1111/bcpt.13537.

Cao, Dong-Sheng, Qing-Song Xu, Qian-Nan Hu, and Yi-Zeng Liang. 2013. "ChemoPy: Freely Available Python Package for Computational Biology and Chemoinformatics." Bioinformatics 29 (8): 1092–94. https://doi.org/10.1093/bioinformatics/btt105.

Caracciolo, Massimo, Pierpaolo Correale, Carmelo Mangano, Giuseppe Foti, Carmela Falcone, Sebastiano Macheda, Maria Cuzzola, et al. 2021. "Efficacy and Effect of Inhaled Adenosine Treatment in Hospitalized COVID-19 Patients." Frontiers in Immunology 12: 734. https://doi.org/10.3389/fimmu.2021.613070.

Chen, Lieyang, Anthony Cruz, Steven Ramsey, Callum J. Dickson, Jose S. Duca, Viktor Hornak, David R. Koes, and Tom Kurtzman. 2019. "Hidden Bias in the DUD-E Dataset Leads to Misleading Performance of Deep Learning in Structure-Based Virtual Screening." PloS One 14 (8): e0220113.

Chen, Lifan, Xiaoqin Tan, Dingyan Wang, Feisheng Zhong, Xiaohong Liu, Tianbiao Yang, Xiaomin Luo, Kaixian Chen, Hualiang Jiang, and Mingyue Zheng. 2020. "TransformerCPI: Improving Compound-Protein Interaction Prediction by Sequence-Based Deep Learning with Self-Attention Mechanism and Label Reversal Experiments." Bioinformatics (Oxford, England) 36 (16): 4406–14. https://doi.org/10.1093/bioinformatics/btaa524.

Chen, Ruolan, Xiangrong Liu, Shuting Jin, Jiawei Lin, and Juan Liu. 2018. "Machine Learning for Drug-Target Interaction Prediction." Molecules 23 (9): 2208. https://doi.org/10.3390/molecules23092208.

Chen, Xing, Chenggang Clarence Yan, Xiaotian Zhang, Xu Zhang, Feng Dai, Jian Yin, and Yongdong Zhang. 2016. "Drug-Target Interaction Prediction: Databases, Web Servers and Computational Models." Briefings in Bioinformatics 17 (4): 696–712. https://doi.org/10.1093/bib/bbv066.

Deganutti, Giuseppe, Filippo Prischi, and Christopher A. Reynolds. 2021. "Supervised Molecular Dynamics for Exploring the Druggability of the SARS-CoV-2 Spike Protein." Journal of Computer-Aided Molecular Design 35 (2): 195–207. https://doi.org/10.1007/s10822-020-00356-4.

Ding, Hao, Ichigaku Takigawa, Hiroshi Mamitsuka, and Shanfeng Zhu. 2014. "Similarity-Based Machine Learning Methods for Predicting Drug-Target Interactions: A Brief Review." Briefings in Bioinformatics 15 (5): 734–47. https://doi.org/10.1093/bib/bbt056.

Doyle, Ken. 2020. "New Uses for Old Drugs: Remdesivir and COVID-19." Promega Connections. July 27, 2020. https://www.promegaconnections.com/new-uses-for-old-drugs-remdesivir-and-covid-19/.

Ferreira, Juliana C., Samar Fadl, Metehan Ilter, Hanife Pekel, Rachid Rezgui, Ozge Sensoy, and Wael M. Rabeh. 2021. "Dimethyl Sulfoxide Reduces the Stability but Enhances Catalytic Activity of the Main SARS-CoV-2 Protease 3CLpro." The FASEB Journal 35 (8): e21774. https://doi.org/10.1096/fj.202100994.

Geiger, Jonathan D., Nabab Khan, Madhuvika Murugan, and Detlev Boison. 2020. "Possible Role of Adenosine in COVID-19 Pathogenesis and Therapeutic Opportunities." Frontiers in Pharmacology 11: 1901. https://doi.org/10.3389/fphar.2020.594487.

Gilson, Michael K., Tiqing Liu, Michael Baitaluk, George Nicola, Linda Hwang, and Jenny Chong. 2016. "BindingDB in 2015: A Public Database for Medicinal Chemistry, Computational Chemistry and Systems Pharmacology." Nucleic Acids Research 44 (D1): D1045–53. https://doi.org/10.1093/nar/gkv1072.

Gönen, Mehmet. 2012. "Predicting Drug–Target Interactions from Chemical and Genomic Kernels Using Bayesian Matrix Factorization." Bioinformatics 28 (18): 2304–10. https://doi.org/10.1093/bioinformatics/bts360.

Günther, Stefan, Michael Kuhn, Mathias Dunkel, Monica Campillos, Christian Senger, Evangelia Petsalaki, Jessica Ahmed, et al. 2008. "SuperTarget and Matador: Resources for Exploring Drug-Target Relationships." Nucleic Acids Research 36 (suppl_1): D919–22. https://doi.org/10.1093/nar/gkm862.

Hashemian, Seyed MohammadReza, Tayebeh Farhadi, and Ali Akbar Velayati. 2020. "A Review on Remdesivir: A Possible Promising Agent for the Treatment of COVID-19." Drug Design, Development and Therapy 14 (August): 3215–22. https://doi.org/10.2147/DDDT.S261154.

Hashemifar, Somaye, Behnam Neyshabur, Aly A. Khan, and Jinbo Xu. 2018. "Predicting Protein–Protein Interactions through Sequence-Based Deep Learning." Bioinformatics 34 (17): i802–10. https://doi.org/10.1093/bioinformatics/bty573.

Hassan, Ameer E. 2021. "An Observational Cohort Study to Assess N-Acetylglucosamine for COVID-19 Treatment in the Inpatient Setting." Annals of Medicine and Surgery 68 (August): 102574. https://doi.org/10.1016/j.amsu.2021.102574.

Hoang, Ba X., Huy Q. Hoang, and Bo Han. 2020. "Zinc Iodide in Combination with Dimethyl Sulfoxide for Treatment of SARS-CoV-2 and Other Viral Infections." Medical Hypotheses 143 (October): 109866. https://doi.org/10.1016/j.mehy.2020.109866.

Huang, Kexin, Tianfan Fu, Lucas M Glass, Marinka Zitnik, Cao Xiao, and Jimeng Sun. 2020. "DeepPurpose: A Deep Learning Library for Drug–Target Interaction Prediction." Bioinformatics 36 (22–23): 5545–47. https://doi.org/10.1093/bioinformatics/btaa1005.

Huang, Ying, Beifang Niu, Ying Gao, Limin Fu, and Weizhong Li. 2010. "CD-HIT Suite: A Web Server for Clustering and Comparing Biological Sequences." Bioinformatics (Oxford, England) 26 (5): 680–82. https://doi.org/10.1093/bioinformatics/btq003.

Jaberi-Douraki, Majid, Emma Meyer, Jim Riviere, Nuwan Indika Millagaha Gedara, Jessica Kawakami, Gerald J. Wyckoff, and Xuan Xu. 2021. "Pulmonary Adverse Drug Event Data in Hypertension with Implications on COVID-19 Morbidity." Scientific Reports 11 (1): 13349. https://doi.org/10.1038/s41598-021-92734-7.

Jacob, Laurent, and Jean-Philippe Vert. 2008. "Protein-Ligand Interaction Prediction: An Improved Chemogenomics Approach." Bioinformatics (Oxford, England) 24 (19): 2149–56. https://doi.org/10.1093/bioinformatics/btn409.

Kim, Jason, Jenny Zhang, Yoonjeong Cha, Sarah Kolitz, Jason Funt, Renan Escalante Chong, Scott Barrett, Rebecca Kusko, Ben Zeskind, and Howard

Kaufman. 2020. "Advanced Bioinformatics Rapidly Identifies Existing Therapeutics for Patients with Coronavirus Disease-2019 (COVID-19)." Journal of Translational Medicine 18. https://doi.org/10.1186/s12967-020-02430-9.

Kozak, Kathryn E., Linda Ouyang, Andriy Derkach, Alexandra Sherman, Susan J. McCall, Christopher Famulare, Jordan Chervin, et al. 2021. "Serum Antibody Response in Patients with Philadelphia-Chromosome Positive or Negative Myeloproliferative Neoplasms Following Vaccination with SARS-CoV-2 Spike Protein Messenger RNA (MRNA) Vaccines." Leukemia, November, 1–3. https://doi.org/10.1038/s41375-021-01457-1.

Laarhoven, Twan van, Sander B. Nabuurs, and Elena Marchiori. 2011. "Gaussian Interaction Profile Kernels for Predicting Drug-Target Interaction." Bioinformatics (Oxford, England) 27 (21): 3036–43. https://doi.org/10.1093/bioinformatics/btr500.

Lana, José Fábio Santos Duarte, Anna Vitória Santos Duarte Lana, Quézia Souza Rodrigues, Gabriel Silva Santos, Riya Navani, Annu Navani, Lucas Furtado da Fonseca, et al. 2021. "Nebulization of Glutathione and N-Acetylcysteine as an Adjuvant Therapy for COVID-19 Onset." Advances in Redox Research 3 (December): 100015. https://doi.org/10.1016/j.arres.2021.100015.

Lee, Heesu, and Jae Wook Lee. 2016. "Target Identification for Biologically Active Small Molecules Using Chemical Biology Approaches." Archives of Pharmacal Research 39 (9): 1193–1201. https://doi.org/10.1007/s12272-016-0791-z.

Lim, Sangsoo, Yijingxiu Lu, Chang Yun Cho, Inyoung Sung, Jungwoo Kim, Youngkuk Kim, Sungjoon Park, and Sun Kim. 2021. "A Review on Compound-Protein Interaction Prediction Methods: Data, Format, Representation and Model." Computational and Structural Biotechnology Journal 19 (January): 1541–56. https://doi.org/10.1016/j.csbj.2021.03.004.

Liu, Hui, Jianjiang Sun, Jihong Guan, Jie Zheng, and Shuigeng Zhou. 2015. "Improving Compound–Protein Interaction Prediction by Building up Highly Credible Negative Samples." Bioinformatics 31 (12): i221–29. https://doi.org/10.1093/bioinformatics/btv256.

M. Veselinovic, Aleksandar, Jovana B. Veselinovic, Jelena V. Zivkovic, and Goran M. Nikolic. 2015. "Application of SMILES Notation Based Optimal Descriptors in Drug Discovery and Design." Current Topics in Medicinal Chemistry 15 (18): 1768–79.

Mazandu, Gaston K., Emile R. Chimusa, Kayleigh Rutherford, Elsa-Gayle Zekeng, Zoe Z. Gebremariam, Maryam Y. Onifade, and Nicola J. Mulder. 2018. "Large-Scale Data-Driven Integrative Framework for Extracting Essential Targets and Processes from Disease-Associated Gene Data Sets." Briefings in Bioinformatics 19 (6): 1141–52. https://doi.org/10.1093/bib/bbx052.

Minhas, Fayyaz ul Amir Afsar, Brian J. Geiss, and Asa Ben-Hur. 2014. "PAIRpred: Partner-Specific Prediction of Interacting Residues from Sequence and Structure." Proteins 82 (7): 1142–55. https://doi.org/10.1002/prot.24479.

Murugan, Natarajan Arul, Sanjiv Kumar, Jeyaraman Jeyakanthan, and Vaibhav Srivastava. 2020. "Searching for Target-Specific and Multi-Targeting Organics for Covid-19 in the Drugbank Database with a Double Scoring Approach." Scientific Reports 10 (1): 19125. https://doi.org/10.1038/s41598-020-75762-7.

Mysinger, Michael M., Michael Carchia, John. J. Irwin, and Brian K. Shoichet. 2012. "Directory of Useful Decoys, Enhanced (DUD-E): Better Ligands and Decoys for Better Benchmarking." Journal of Medicinal Chemistry 55 (14): 6582–94. https://doi.org/10.1021/jm300687e.

Nguyen, Thin, Hang Le, Thomas P Quinn, Tri Nguyen, Thuc Duy Le, and Svetha Venkatesh. 2021. "GraphDTA: Predicting Drug–Target Binding Affinity with Graph Neural Networks." Bioinformatics 37 (8): 1140–47. https://doi.org/10.1093/bioinformatics/btaa921.

Öztürk, Hakime, Arzucan Özgür, and Elif Ozkirimli. 2018. "DeepDTA: Deep Drug–Target Binding Affinity Prediction." Bioinformatics 34 (17): i821–29.

Öztürk, Hakime, Elif Ozkirimli, and Arzucan Özgür. 2019. "WideDTA: Prediction of Drug-Target Binding Affinity." ArXiv:1902.04166 [Cs, q-Bio, Stat], February. http://arxiv.org/abs/1902.04166.

Paul, Steven M., Daniel S. Mytelka, Christopher T. Dunwiddie, Charles C. Persinger, Bernard H. Munos, Stacy R. Lindborg, and Aaron L. Schacht. 2010. "How to Improve R&D Productivity: The Pharmaceutical Industry's Grand Challenge." Nature Reviews Drug Discovery 9 (3): 203–14. https://doi.org/10.1038/nrd3078.

Riley, Patrick. 2019. "Three Pitfalls to Avoid in Machine Learning." Nature 572 (7767): 27–29. https://doi.org/10.1038/d41586-019-02307-y.

Rohrer, Sebastian G., and Knut Baumann. 2009. "Maximum Unbiased Validation (MUV) Data Sets for Virtual Screening Based on PubChem Bioactivity Data." Journal of Chemical Information and Modeling 49 (2): 169–84. https://doi.org/10.1021/ci8002649.

Schirle, Markus, and Jeremy L. Jenkins. 2016. "Identifying Compound Efficacy Targets in Phenotypic Drug Discovery." Drug Discovery Today 21 (1): 82–89. https://doi.org/10.1016/j.drudis.2015.08.001.

Senger, Mario Roberto, Tereza Cristina Santos Evangelista, Rafael Ferreira Dantas, Marcos Vinicius da Silva Santana, Luiz Carlos Saramago Gonçalves, Lauro Ribeiro de Souza Neto, Sabrina Baptista Ferreira, and Floriano Paes Silva-Junior. 2020. "COVID-19: Molecular Targets, Drug Repurposing and New Avenues for Drug Discovery." Memórias Do Instituto Oswaldo Cruz 115 (October). https://doi.org/10.1590/0074-02760200254.

Sieg, Jochen, Florian Flachsenberg, and Matthias Rarey. 2019. "In Need of Bias Control: Evaluating Chemical Data for Machine Learning in Structure-Based Virtual Screening." Journal of Chemical Information and Modeling 59 (3): 947–61. https://doi.org/10.1021/acs.jcim.8b00712.

Singh, Gagandeep, Vishal Srivastava, Ritpratik Mishra, Gaurav Goel, and Tapan Chaudhuri. 2020. "Old Arsenal to Combat New Enemy: Repurposing of Commercially Available FDA Approved Drugs Against Main Protease of SARS-CoV2.," October. https://doi.org/10.26434/chemrxiv.13032578.v1.

Siramshetty, Vishal B, Oliver Andreas Eckert, Björn-Oliver Gohlke, Andrean Goede, Qiaofeng Chen, Prashanth Devarakonda, Saskia Preissner, and Robert Preissner. 2018. "SuperDRUG2: A One Stop Resource for Approved/Marketed Drugs." Nucleic Acids Research 46 (D1): D1137–43. https://doi.org/10.1093/nar/gkx1088.

Thafar, Maha, Arwa Bin Raies, Somayah Albaradei, Magbubah Essack, and Vladimir B. Bajic. 2019. "Comparison Study of Computational Prediction Tools for Drug-Target Binding Affinities." Frontiers in Chemistry 7: 782. https://doi.org/10.3389/fchem.2019.00782.

Tsubaki, Masashi, Kentaro Tomii, and Jun Sese. 2019. "Compound-Protein Interaction Prediction with End-to-End Learning of Neural Networks for Graphs and Sequences." Bioinformatics (Oxford, England) 35 (2): 309–18. https://doi.org/10.1093/bioinformatics/bty535.

Vapnik, Vladimir, and Rauf Izmailov. 2017. "Knowledge Transfer in SVM and Neural Networks." Annals of Mathematics and Artificial Intelligence, February, 1–17. https://doi.org/10.1007/s10472-017-9538-x.

Wang, Pei-Gang, Dong-Jiang Tang, Zhan Hua, Zai Wang, and Jing An. 2020. "Sunitinib Reduces the Infection of SARS-CoV, MERS-CoV and SARS-CoV-2 Partially by Inhibiting AP2M1 Phosphorylation." Cell Discovery 6 (1): 1–5. https://doi.org/10.1038/s41421-020-00217-2.

Wang, Shudong, Zhenzhen Du, Mao Ding, Renteng Zhao, Alfonso Rodriguez-Paton, and Tao Song. 2020. "LDCNN-DTI: A Novel Light Deep Convolutional Neural Network for Drug-Target Interaction Predictions." In 2020 IEEE International Conference on Bioinformatics and Biomedicine (BIBM), 1132–36. https://doi.org/10.1109/BIBM49941.2020.9313585.

Wang, Yuhao, and Jianyang Zeng. 2013. "Predicting Drug-Target Interactions Using Restricted Boltzmann Machines." Bioinformatics 29 (13): i126–34. https://doi.org/10.1093/bioinformatics/btt234.

Weisberg, Ellen, Alexander Parent, Priscilla L. Yang, Martin Sattler, Qingsong Liu, Qingwang Liu, Jinhua Wang, et al. 2020. "Repurposing of Kinase Inhibitors for Treatment of COVID-19." Pharmaceutical Research 37 (9): 167. https://doi.org/10.1007/s11095-020-02851-7.

Wishart, David S., Craig Knox, An Chi Guo, Dean Cheng, Savita Shrivastava, Dan Tzur, Bijaya Gautam, and Murtaza Hassanali. 2008. "DrugBank: A Knowledgebase for Drugs, Drug Actions and Drug Targets." Nucleic Acids Research 36 (suppl_1): D901–6. https://doi.org/10.1093/nar/gkm958.

Xu, Jimin, Yu Xue, Richard Zhou, Pei-Yong Shi, Hongmin Li, and Jia Zhou. 2021. "Drug Repurposing Approach to Combating Coronavirus: Potential Drugs and Drug Targets." Medicinal Research Reviews 41 (3): 1375–1426. https://doi.org/10.1002/med.21763.

Zhang, Weilin, Jianfeng Pei, and Luhua Lai. 2017. "Computational Multitarget Drug Design." Journal of Chemical Information and Modeling 57 (3): 403–12. https://doi.org/10.1021/acs.jcim.6b00491.

Zhang, Xiao-Meng, Li Liang, Lin Liu, and Ming-Jing Tang. 2021. "Graph Neural Networks and Their Current Applications in Bioinformatics." Frontiers in Genetics 12: 1073. https://doi.org/10.3389/fgene.2021.690049.

Zhao, Tianyi, Yang Hu, Linda R. Valsdottir, Tianyi Zang, and Jiajie Peng. 2021. "Identifying Drug-Target Interactions Based on Graph Convolutional Network and Deep Neural Network." Briefings in Bioinformatics 22 (2): 2141–50. https://doi.org/10.1093/bib/bbaa044.